\documentclass[traditabstract]{aa}
\usepackage{graphicx}
\usepackage[varg]{txfonts}
\usepackage{natbib}
\usepackage{url}
\usepackage{color}
\bibpunct{(}{)}{;}{a}{}{,} 
\newcommand\teff{T_{\rm eff}}

\usepackage{hyperref}
\hypersetup{
     breaklinks,
     colorlinks   = true,
     citecolor    = blue,
     urlcolor     = blue,
     linkcolor    = blue
}

\begin{document}

\title{
Chemical evolution of the Galactic bulge as traced\\ by microlensed dwarf and subgiant stars
}
\subtitle{
VII. Lithium
\thanks{Based on data obtained with the European Southern Observatory telescopes (Proposal ID:s 87.B-0600, 88.B-0349, 89.B-0047, 90.B-0204, 91.B-0289, 92.B-0626, 93.B-0700), and the Magellan Clay telescope at the Las Campanas observatory.}
\fnmsep
\thanks{Table~1 is available in electronic form at the CDS via
anonymous ftp to {\tt cdsarc.u-strasbg.fr (130.79.128.5)} or via
\url{http://cdsweb.u-strasbg.fr/cgi-bin/qcat?J/A+A/XXX/AXX}.
}
}
\titlerunning{Chemical evolution of the Galactic bulge as traced by microlensed dwarf and subgiant stars. VII. Lithium}

\author{
T.~Bensby\inst{1}
\and
S.~Feltzing\inst{1}
\and
J.C.~Yee\inst{2}
\and
J.A.~Johnson\inst{3}
\and
A.~Gould\inst{3,4,5}
\and
M.~Asplund\inst{6}
\and
J.~Mel\'endez\inst{7}
\and
S.~Lucatello\inst{8}
 }

\institute{Lund Observatory, Department of Astronomy and
Theoretical physics,
Box 43, SE-221\,00 Lund, Sweden \\
\email{tbensby@astro.lu.se}
\and
Center for Astrophysics $|$ Harvard \& Smithsonian, 60 Garden St., Cambridge, MA 02138, USA
\and
Department of Astronomy, Ohio State University, 140 W. 18th Avenue,
Columbus, OH 43210, USA
\and
Max Planck Institute for Astronomy, K\"onigstuhl 17, 69117 Heidelberg, Germany
\and
Korea Astronomy and Space Science Institute Institute, 305-348 Daejon, Republic of Korea
\and
Research School of Astronomy \& Astrophysics, Mount Stromlo Observatory, Cotter Road,
Weston Creek, ACT 2611, Australia
\and
Departamento de Astronomia do IAG/USP, Universidade de S\~ao Paulo,
Rua do Mat\~ao 1226, S\~ao Paulo, 05508-900, SP, Brasil
\and
INAF-Astronomical Observatory of Padova, Vicolo dell'Osservatorio 5,
35122 Padova, Italy
}


\date{Received 23 December 2019 / Accepted XX Xxxxxxx 202X}

 \abstract{
Lithium abundances are presented for 91 dwarf and subgiant stars in the Galactic bulge. The analysis is based on line synthesis of the $^7$Li line at 6707\,{\AA} in high-resolution spectra obtained during gravitational microlensing events, when the brightnesses of the targets were highly magnified. Our main finding is that the bulge stars at sub-solar metallicities, and that are older than about eight billion years, does not show any sign of Li production, that is, the Li trend with metallicity is flat (or even slightly declining). This indicates that no lithium was produced during the first few billion years in the history of the bulge. This finding is essentially identical to what is seen for the (old) thick disk stars in the Solar neighbourhood, and adds  another piece of evidence for a tight connection between the metal-poor bulge and the Galactic thick disk. For the bulge stars younger than about eight billion years, the sample contains a group of stars at very high metallicities at $\rm [Fe/H]\approx +0.4$ that have lithium abundances in the range $\rm A(Li)=2.6-2.8$.  In the Solar neighbourhood the lithium abundances have been found to peak at a $\rm A(Li)\approx 3.3$ at $\rm [Fe/H]\approx +0.1$ and then decrease by $0.4-0.5$\,dex when reaching $\rm [Fe/H]\approx+0.4$. The few bulge stars that we have at these metallicities, seem to support this declining A(Li) trend. This could indeed support the recent claim that the low A(Li) abundances at the highest metallicities seen in the Solar neighbourhood could be due to stars from the inner disk, or the bulge region, that have migrated to the Solar neighbourhood.
}
   \keywords{
   Gravitational lensing: micro ---
   Galaxy: bulge ---
   Galaxy: formation ---
   Galaxy: evolution ---
   Stars: abundances
   }
   \maketitle

\section{Introduction}
\label{sec:introduction}

The Li abundance trends in the Milky Way disk and halo have been extensively mapped during the last two decades by high-resolution spectroscopic studies of nearby stars \citep[e.g][]{chen2001,lambert2004,ghezzi2010li,ramirez2012,delgadomena2014,guiglion2016,pavlenko2018,bensby2018}. The collective consensus is that the amount of Li observed in metal-poor halo stars lies at a level of about $\rm A(Li)\approx2.2-2.3$, and that it remains essentially at this level over a wide range of metallicities up to at least $\rm [Fe/H]\approx -1$ \citep{spite1982}. At the very lowest metallicities below $\rm [Fe/H]\lesssim -4$ it shows a decreasing trend \citep[e.g.][]{sbordone2010,bonifacio2018}. At higher metallicities (greater than about $\rm [Fe/H]\gtrsim-1$) where the Galactic disk(s) dominate, the  Spite plateau extends to about $\rm [Fe/H]\approx-0.6$, after which the Li trend shows a steep increase, peaking at $\rm A(Li)\approx3.2$ around, or slightly above, solar metallicities. However, if only (thick) disk stars with ages greater than about eight billion years are considered the Li trend above $\rm [Fe/H]=-1$ remains flat, or slightly declining, towards solar metallicities \citep{bensby2018}. On the other hand, if (thin) disk stars younger than about eight billion years are considered, the increasing Li trend towards and beyond solar metallicities emerges. At the very highest metallicities ($\rm [Fe/H]\gtrsim+0.2$) the Li trend again shows a tendency of decreasing with [Fe/H] \citep{delgadomena2015,guiglion2016,fu2018,bensby2018}.

This varying Li trend with metallicity in different stellar populations show that Li is not only produced in stars but also depleted. While the depletion process is fairly well-understood the production sites are widely debated. Small amounts of Li were made in the Big Bang nucleosynthesis, theoretically putting it at a level of 2.6-2.7 \citep[e.g.][]{cyburt2016}. This is in contrast with the observations on the Spite plateau observed in metal-poor stars that is about 0.4\,dex lower, and is often considered as the primordial value. This cosmological Li problem remains unsolved \citep[e.g.][]{melendez2010li,sbordone2010}.

The non-increasing Li trend in the local thick disk means that no Li was produced during the thick disk era that took place during the first few billion years of the Milky Way. Only later, in the thin disk, the Li trend increases strongly signalling the first production of Li since the Big Bang. Possible Li production sites include for example supernovae type II, cosmic rays, AGB stars, RGB stars, and novae \citep[e.g.][]{dantona1991,romano1999}. \cite{romano2001} explored the contribution of the different sources in their two-infall chemical evolution model of the Milky Way and found that the most important contributors were low-mass red giants, and that novae were needed in order to reproduce the steep rise in the Li trend at higher metallicities. This has recently been further investigated by \cite{cescutti2019} that indeed confirms that novae most likely are responsible for the production of the great majority of the Galactic Li \citep[see also][]{matteucci2010}.
At the same time \cite{cescutti2019} show that core-collapse supernovae, RGB, and AGB stars are less significant Li production sites, and can to a large degree be neglected compared to the amounts of Li being produced in novae and to some degree in cosmic ray spallation processes.

The recent discovery of a decreasing Li trend at the very highest metallicities is an unexpected result and several explanations have been proposed. First, maybe metal-rich stars are less efficient in producing Li. However, the current models of stellar evolution and nucleosynthesis does not support such a scenario as it means that  Li that is being produced in stars is starting to decline during the last few billion years, and would be the first element showing this behaviour \citep{prantzos2017}. Another possibility could be that the apparent decline in the Li trend in the solar neighbourhood is due to stars that have migrated here from the inner disk region, or even the bulge region, and that they show a lower Li abundance due to a faster star formation history in the inner disk/bulge regions \citep{guiglion2019}. Recently, \cite{grisoni2019} provided a new explanation, in which novae are the main Li producers and the decline at the highest metallicities could be due to a metallicity dependent fraction of the binary systems that are the novae precursors.   No satisfactory explanation has so far been provided as to whether the migration scenario or the metallicity dependent binary fraction scenario is to be preferred.

The observational data that has been considered so far are to a large degree confined to the solar neighbourhood (see references above). Even though large spectroscopic surveys such as the Gaia-ESO survey \citep{gilmore2012} and GALAH \citep{desilva2015} reaching farther out in the disk, and recent results also include Li \citep{fu2018,buder2018}, they do however not reach into the very inner or outer disk regions. Li in the bulge is particularly poorly constrained, and as the bulge is a major structural component of the Milky Way, this hampers our ability to get a complete view of the production and depletion of Li in the Milky Way. Also, an interesting aspect of the bulge is that it contains some of the most metal-rich stars ever observed, reaching $\rm [Fe/H]\gtrsim +0.5$. It could therefore hint about the Li decline seen in local dwarf samples, and if it is also present in the bulge stars, and whether it continues at the even higher metallicities seen in the bulge. The few studies that have measured Li abundances in bulge stars have mainly done so using RGB stars \cite{gonzalez2009,lebzelter2012li} or AGB stars \citep{uttenthaler2007li}. These stars do however not reflect the native Li abundance that they were born with due to the deep convection zones in the outer regions of their atmospheres that will bring down Li to warmer regions where it is destroyed. Li in dwarf stars warmer than about 5800\,K are not affected by this. Dwarf stars in the bulge are however very faint and are very difficult to observe under normal circumstances.

In \cite{bensby2017} we presented detailed elemental abundances and stellar ages for 91 microlensed dwarf and subgiant stars in the Galactic bulge. It was shown that the bulge has very complex age and abundance distributions, and that it most likely consists of stars from the other Galactic stellar populations. In particular it was shown that the metal-poor bulge and the thick disk have very similar abundance trends \citep[see also][]{melendez2008,alvesbrito2010,bensby2011,gonzalez2011,bensby2013,jonsson2017}, that they have similar ages that in general are greater than about 8\,Gyr, and that the metal-rich bulge in some ways resemble what is observed in the local thin disk.

Regarding Li in the bulge, \cite{minniti1998} was using microlensing events to study bulge dwarf stars and claimed a detection of Li in MACHO- 1997-BLG-45/47. Later, \cite{cavallo2003} re-analysed the spectrum obtained by \cite{minniti1998} and could not confirm the Li detection owing to the limited S/N. The first clear detection of Li in an un-evolved dwarf star was presented in \cite{bensby2010li} for the metal-poor bulge dwarf MOA-2010-BLG-285S, and for another five microlensed bulge dwarf stars in \cite{bensby2011}, for which the Li line could be detected. These studies showed that the metal-poor bulge dwarf stars have Li abundances consistent with the Spite plateau, and that the ones more metal-rich have a large spread in Li. In this study we utilise the now larger sample of microlensed bulge dwarf stars to investigate whether the distinct Li trends seen in the solar neighbourhood persist also in the bulge. If so, this would be another clear indication of the connection between the bulge and the disk(s). If possible we also aim to explore Li abundances at the very highest metallicities. The bulge contains the most metal-rich stars ever observed and should be the optimal place to explore whether the declining Li trend with metallicity is also present there.

\section{Abundance analysis}

\begin{figure*}
\centering
\resizebox{\hsize}{!}{
\includegraphics[viewport= 15 0 400 504,clip]{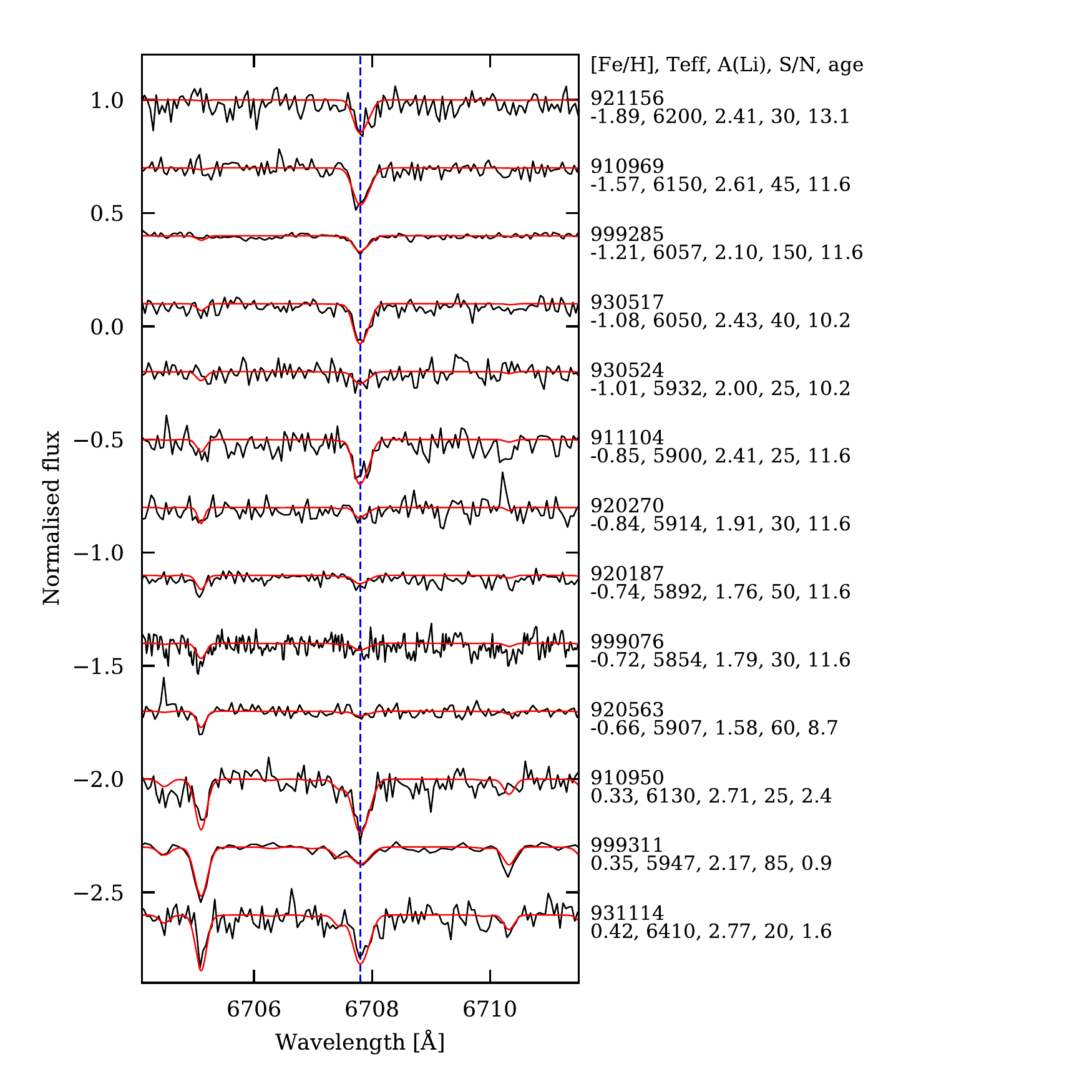}
\includegraphics[viewport= 15 0 400 504,clip]{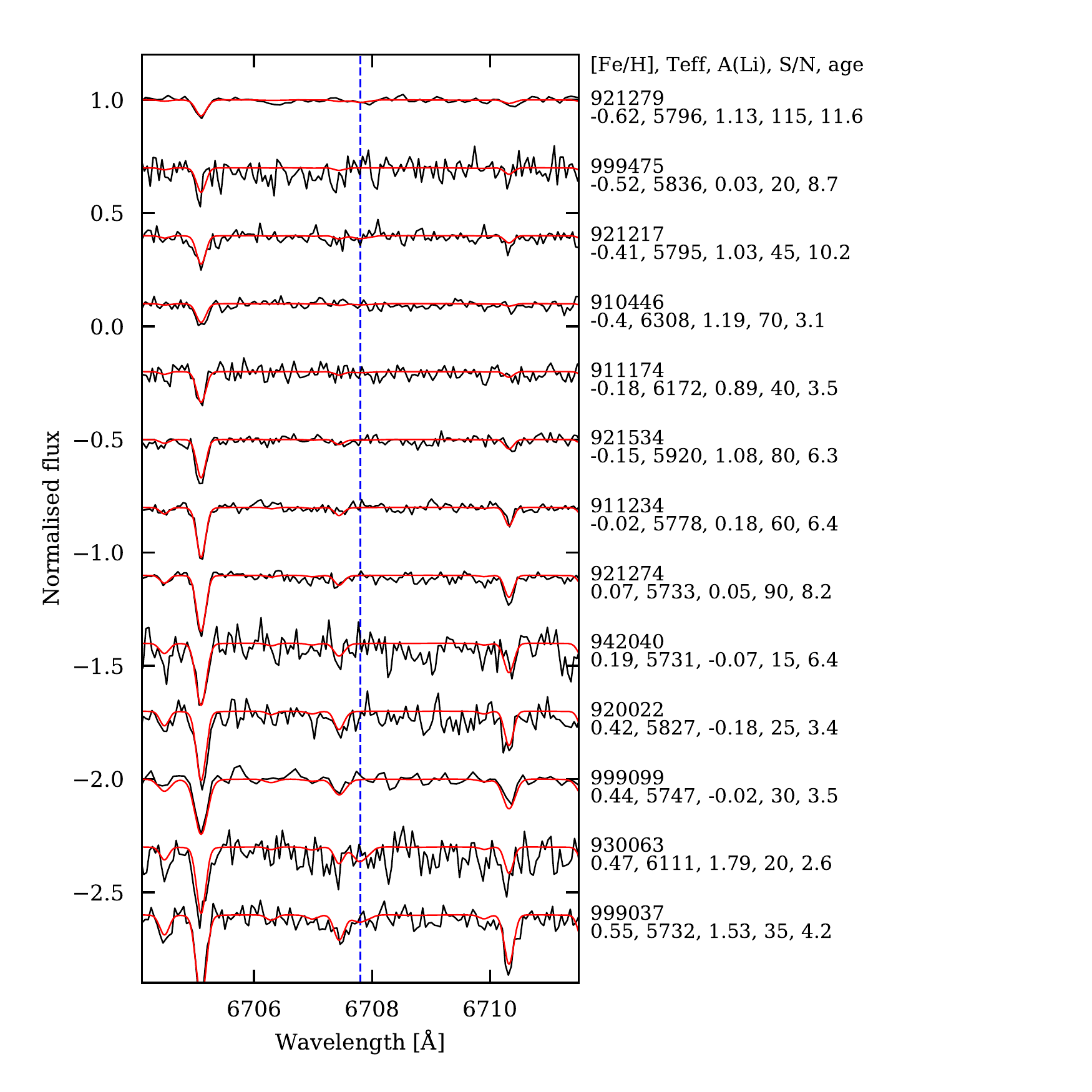}}
\caption{{\sl Left-hand side:} Normalised spectra for 13 stars that have $\teff >5800$\,K and where the Li line at 6707.8\,{\AA} could be detected and the Li abundance measured. The vertical dotted line shows the wavelength where the Li line is located and the red solid lines show the best fitting synthetic spectra. {\sl Right-hand side:} Normalised spectra for 13 stars that have $\teff >5700$\,K and where the Li line at 6707.8\,{\AA} could not be detected, and only A(Li) upper limits could be determined. The vertical dotted line shows the location where the Li line should be. For each spectrum we give [Fe/H], $\teff$, A(Li), the measured S/N per pixel, and the age of the star. 
\label{fig:li_spectra}}
\end{figure*}

We analyse Li for the full sample of 91 microlensed dwarf stars using the same methodology as in \cite{bensby2018} where Li was investigated for the 714 nearby thin and thick disk stars from \cite{bensby2014}. In summary, Li abundances are determined through spectral line synthesis of the $^7$Li line at 670.7\,nm. For the line synthesis we used the Uppsala MARCS model atmospheres \citep{gustafsson2008} together with SME \citep{valenti1996} as a spectral synthesiser. The linelist and atomic data are the same as in \cite{bensby2018}. Input stellar parameters and their uncertainties were adopted from \cite{bensby2017} where the sample is described in detail. Out of the 91 microlensed bulge dwarfs, the Li line could be detected, and Li abundances determined, for 22 stars. For the remaining 69 stars only upper limits on the Li abundance could be estimated. These non-detections are most likely due to the wide variation in the signal-to-noise ratio that the obtained microlensed dwarf star spectra have, meaning that the Li line could not be resolved from the continuum noise in many cases.

We show in Fig.~\ref{fig:li_spectra}, on the left-hand side, the observed and the best fitting synthetic spectra for 13 stars that have $\teff>5800$\,K and that have estimated Li abundances. In most stars the Li line is clearly visible, even in those stars where the $S/N$ is getting low around $S/N=25-30$. Figure~\ref{fig:li_spectra}, right-hand side, shows the observed spectra for the 13 stars that have $\teff>5700$\,K, and for which we could only determine upper limits on the Li abundances.

Li abundances are sensitive to changes in $\teff$, while hardly at all to changes in $\log g$ and [Fe/H]. Random uncertainties were estimated by varying the effective temperatures according to their uncertainties given in \cite{bensby2017}. This gives upper and lower limits to the Li abundances. Li NLTE corrections from \cite{lind2009} were added to the Li abundances. The Li abundances, Li uncertainties, and applied NLTE corrections are given in Table~\ref{tab:abundances}.

Systematic uncertainties are difficult to estimate, but as we will compare our derived Li abundances to the Li abundances for the sample of 714 nearby F and G dwarf stars, for which the exact same methods have been applied, the relative uncertainties should be at a minimum between these studies. \cite{bensby2017} also compared the Li abundances for stars in common with \cite{ramirez2012} and \cite{delgadomena2015} and found good agreement.

\begin{table}
\setlength{\tabcolsep}{1.0mm}
\centering
\caption{Li abundances for the microlensed bulge dwarfs\tablefootmark{$\dagger$}
}
\footnotesize
\label{tab:abundances}
\begin{tabular}{ccccccc}
\hline\hline
\noalign{\smallskip}
Name   &
Nr &
A(Li) &
A(Li)$_{l}$ &
A(Li)$_{h}$ &
$\Delta_{\rm NLTE}$ &
Flag \\
\noalign{\smallskip}
\hline
\noalign{\smallskip}
  \vdots & \vdots & \vdots & \vdots & \vdots & \vdots & \vdots \\
 MOA-2010-BLG-167S & 910167 & 1.27 & 1.22 & 1.31 & 0.08 & 0 \\
  \vdots & \vdots & \vdots & \vdots & \vdots & \vdots & \vdots  \\
\noalign{\smallskip}
\hline
\end{tabular}
\tablefoot{
\tablefoottext{$\dagger$}{
For each star we give the NLTE corrected A(Li) abundance and the low and high values determined by changing the effective temperatures by their uncertainties. The Li NLTE corrections that were added are also given. The last column is a flag where the value "0" means that it is a value from a well-fitted line. A value of "1" means that the Li abundance is an upper limit.
All other parameters are given in \cite{bensby2017} where the two identifiers (Name and Nr) are given in the same way for easy cross-matching.
This table is only available in electronic form at the CDS via anonymous ftp to
\url{cdsarc.u-strasbg.fr (130.79.128.5)} or via
\url{http://cdsweb.u-strasbg.fr/cgi-bin/qcat?J/A+A/XXX/AXX}
}}
\end{table}
                   
\section{Results}

\subsection{Li and effective temperature}

\begin{figure}
\centering
\resizebox{\hsize}{!}{
\includegraphics[viewport= 0 0 504 485,clip]{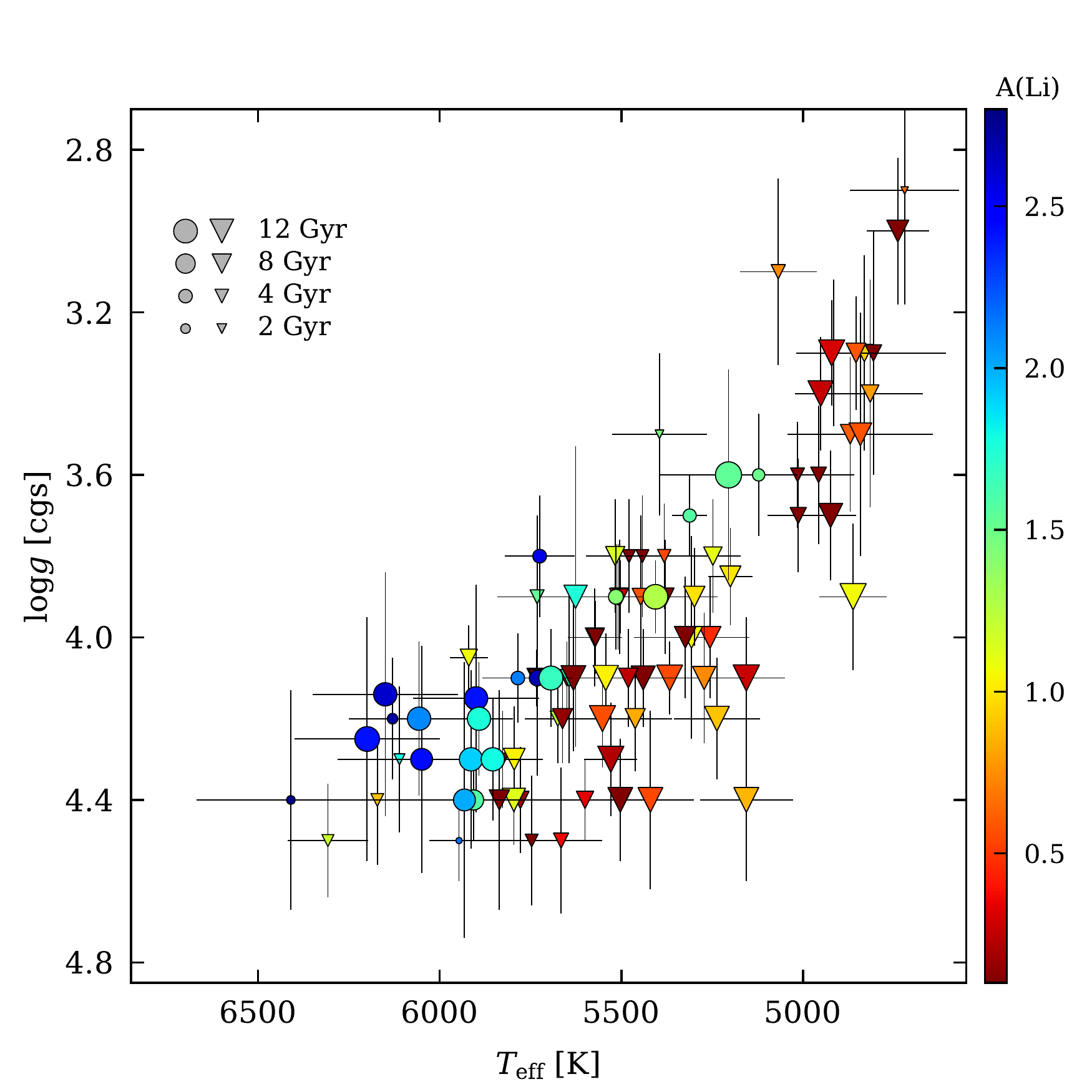}}
\caption{HR diagram for the microlensed dwarf sample. The stars have been colour-coded based on the measured Li abundances (according to the colour-bar on the right-hand side).  Triangles represent those stars where only upper abundance limits could be determined. The sizes of the markers are scaled with the estimated ages of the stars.
\label{fig:li_hr}}
\end{figure}

\begin{figure}
\centering
\resizebox{\hsize}{!}{
\includegraphics[viewport= 0 0 504 360,clip]{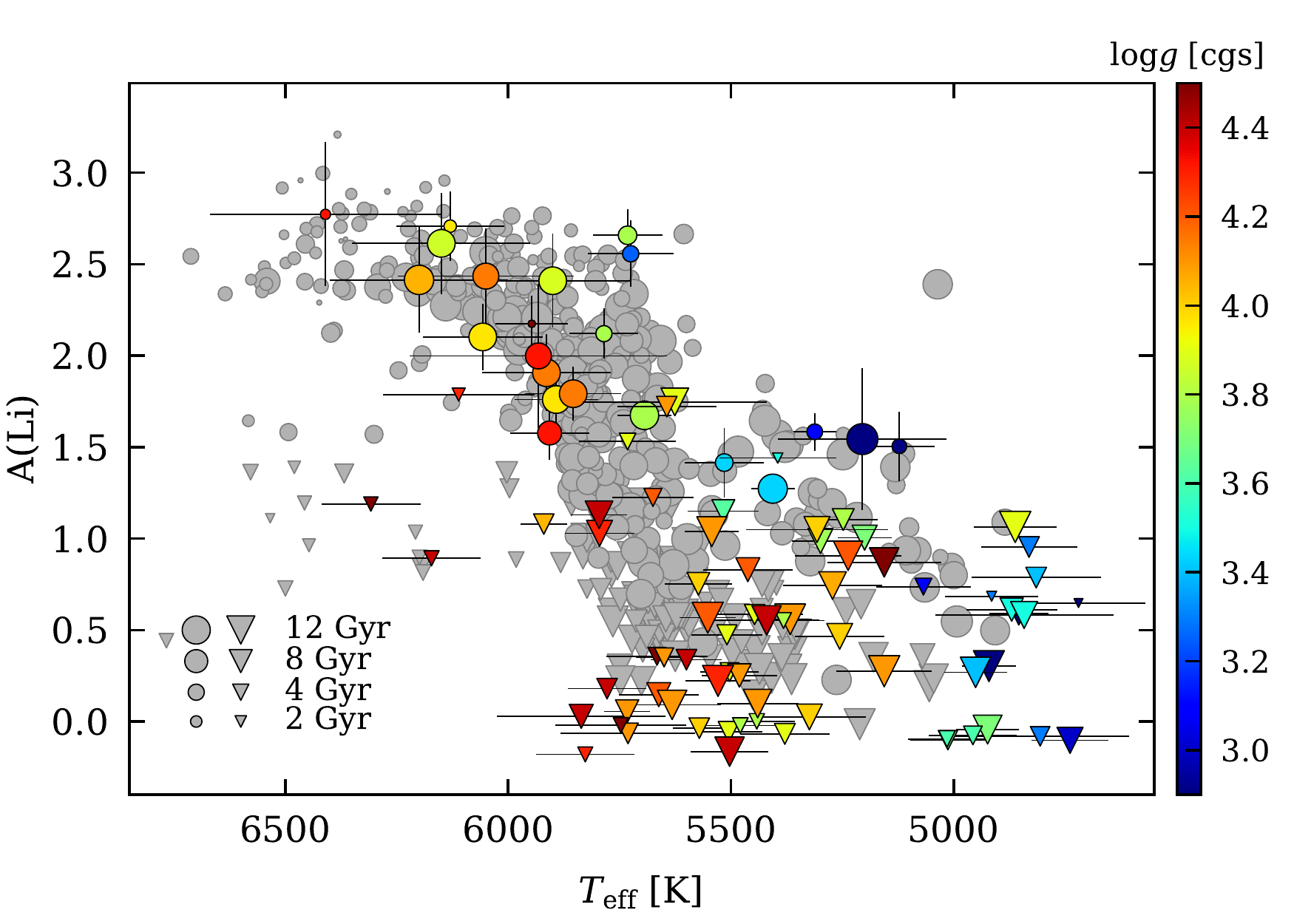}}
\caption{Li abundances versus $\teff$ for the microlensed dwarf sample.   Triangles represent those stars where we could only determine upper limits on the abundances. The grey circles and triangles in the background represent the 714 stars from \cite{bensby2014} for which Li abundances were determined in \cite{bensby2018} in the exact same way as for the microlensed bulge dwarfs. The sizes of the markers have been scaled with the estimated ages, and the bulge stars have also been colour-coded based on their surface gravities according to the colour-bar on the right-hand side.
\label{fig:li_teff}}
\end{figure}

Figure~\ref{fig:li_hr} shows an HR diagram for the microlensed bulge stars. The highest Li abundances are, as expected, seen for the stars with the highest temperatures in the turn-off region. This is in agreement with what was found in \cite{bensby2018}, that use the exact same methods as in this study, as well as in other studies (see references in Sect.~\ref{sec:introduction}). The microlensed bulge stars also show the general increase in Li abundance with effective temperature that has been seen in most studies of Li. This is shown in Fig.~\ref{fig:li_teff}, and it is evident that the bulge dwarfs follow the same pattern as outlined by the sample of 714 nearby disk stars from \cite{bensby2018}. The decline with decreasing temperature is a result of Li being destroyed in stars cooler than about 5800\,K. To trace the Galactic chemical evolution of Li one therefore needs to consider stars with sufficiently high temperature so that the observed surface abundance of Li not has been affected by processes internal to the star. The sample contains 19 stars with effective temperatures higher than 5800\,K, and six of those have only upper limit estimates of the Li abundances.

There are three bulge stars with $6100\lesssim\teff\lesssim6300$\,K and that have Li upper limits significantly below $\rm A(Li)<2.0$ (star no. 910446 with $\teff=6308$\,K and $\rm A(Li)=1.19$, star no. 911174 with $\teff=6172$\,K and $\rm A(Li)=0.89$, and star no. 930063 with $\teff=6111$\,K and $\rm A(Li)=1.79$, see Fig.~\ref{fig:li_spectra}). Similar stars can also seen in the \cite{bensby2018} sample (see Fig.~\ref{fig:li_teff}). The explanation for the low Li abundances in these stars is uncertain. They are not hot enough to be part of the dip-like feature that appears in A(Li) versus $\teff$ , centred around 6700 K \citep{boesgaard2016}, and that was first seen in open clusters \citep{boesgaard1986}. On the other hand there is a possibility that they could be blue straggler stars, that observationally are well established to show essentially no measurable Li \citep[e.g.][]{glaspey1994,carney2005b}.  For instance, a few hot halo dwarf stars that are ultra-Li-depleted has been suggested to be blue stragglers \citep{ryan2001b}, and was recently confirmed by \cite{bonifacio2019}. While the exact formation scenario of blue stragglers is not settled blue stragglers formed by binary star coalescence show a very strong Li depletion (more than 5\,dex), while collisional blue stragglers show some remaining Li at the level $\rm A(Li) \approx 1.1$ \citep{glebbeek2010}, which agrees reasonably with the upper limit Li abundances for two of the three bulge stars (bulge star no. 910446 and 911174). The metallicities of these two stars are $\rm [Fe/H]=-0.41$, and $\rm -0.18$, respectively. If these two stars indeed were to be considered as blue straggler stars, their stellar age estimates could be too young. However, in the case of the solar analog HIP 10725, that has several signatures of blue stragglers (non detectable lithium, non detectable beryllium, large excess of neutron capture elements, and more), just a small amount of mass transferred (less than 0.01\,M$_\odot$) was enough to introduce the observed excess of n-capture elements, and due to the extra-mixing because of the transfer of angular momentum, lithium (and beryllium) were depleted \citep{schirbel2015}. So, even if there are signatures of a blue straggler phenomenon, the age may not change much, except of course in more dramatic cases such as a star merger or significant mass transfer. In any case, the presence or non-presence of blue stragglers in our sample will not affect our conclusions regarding the Galactic evolution of Li as it is based on stars where Li is readily detected and measured.

\subsection{Li and metallicity}

\begin{figure}
\centering
\resizebox{\hsize}{!}{
\includegraphics[viewport= 0 0 504 360,clip]{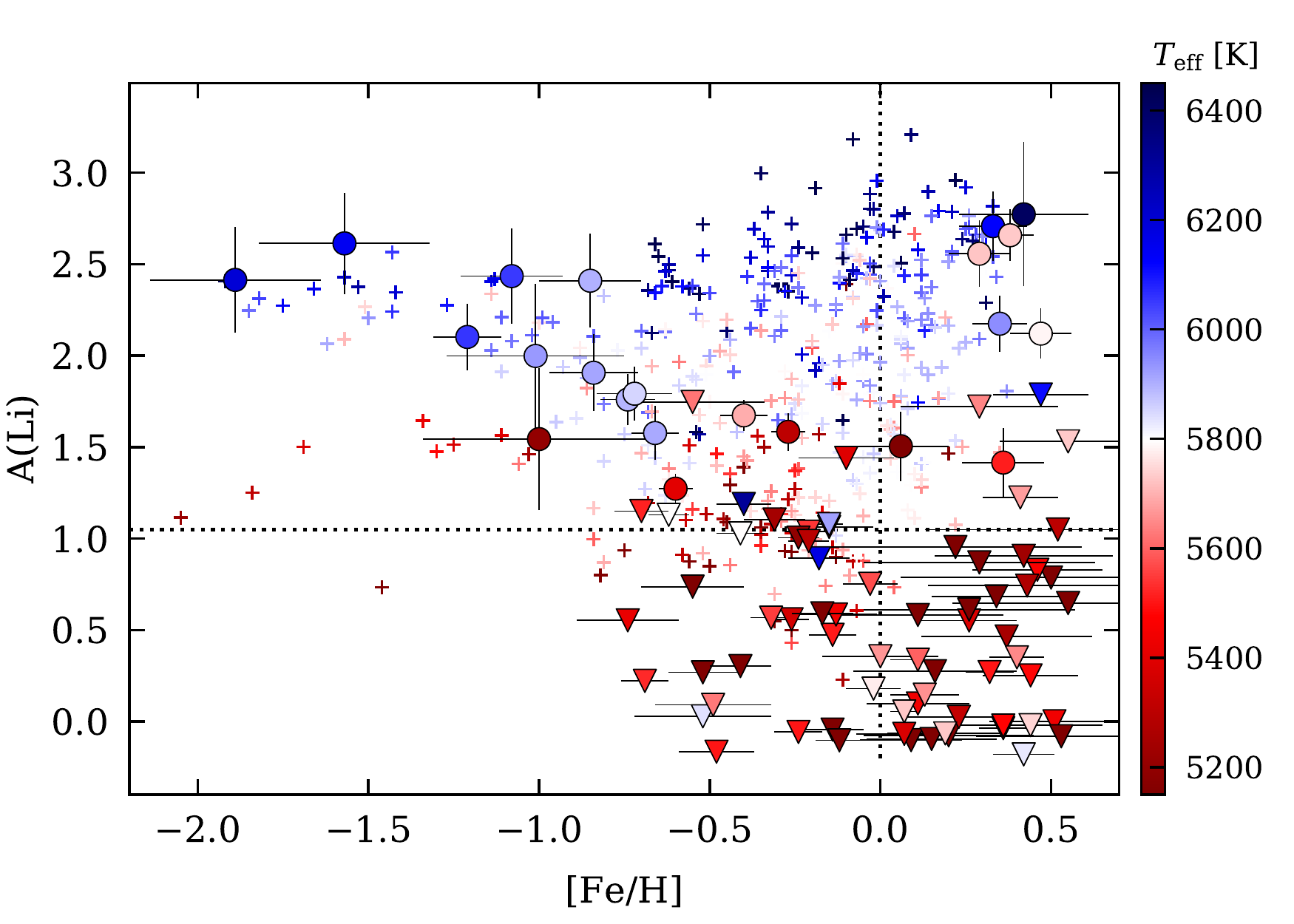}}
\caption{Li abundances versus $\teff$ for the microlensed dwarf sample. The stars have been colour-coded based on the effective temperatures of the stars.  Triangles represent those stars where we could only determine upper limits on the abundances. The grey-coloured plus-signs in the background represent the 714 stars from \cite{bensby2014} for which Li abundances were determined in \cite{bensby2018} in the exact same way as for the microlensed bulge dwarfs.
The dotted lines show the solar values with the solar photospheric Li abundance taken from \cite{asplund2009}.
\label{fig:li_feh}}
\end{figure}

Figure~\ref{fig:li_feh} shows the results in the A(Li)-[Fe/H] plane with the nearby disk stars from \cite{bensby2018}, following the same colour-coding, in the background. There are ten bulge stars at $\rm |Fe/H]\lesssim -0.6$ that have $\teff>5800$\,K and at the metal-poor end, within the uncertainties, these follow the Spite plateau as here also outlined by the disk stars in the background. The A(Li) trend appears to decrease to lower A(Li) values with [Fe/H] as one approaches $\rm [Fe/H]\approx -0.6$. Considering the six stars that have $\teff>5800$\,K and upper limit A(Li) in the metallicity range $\rm -0.4\lesssim [Fe/H]\lesssim-0.2$ the declining A(Li) trend might continue to even higher metallicities.

At super-solar metallicities the sample only contains three stars with $\teff>5800$\,K and that at the same time have well-determined Li abundances. They are grouped at the very highest metallicities around $\rm [Fe/H]\approx 0.3-0.4$ and they follow the general A(Li) trend as outlined by the nearby disk stars. There are also two stars with $\teff>5800$\,K at these high metallicities that have upper limit measurements at even lower A(Li) values - one at $\rm A(Li)\approx1.6$ and one at  $\rm A(Li)\approx-0.2$. In the metallicity range $\rm -0.6\lesssim [Fe/H]\lesssim+0.3$ the sample only contains a few stars with well-determined Li abundances, and all have temperatures below 5800\,K. The stars with $\teff>5800$\,K in this metallicity range have upper limits on the Li abundances, meaning that it will be difficult to reliably trace how Li has evolved at higher metallicities.

\begin{figure*}
\centering
\resizebox{\hsize}{!}{
\includegraphics[viewport= 0 0 504 360,clip]{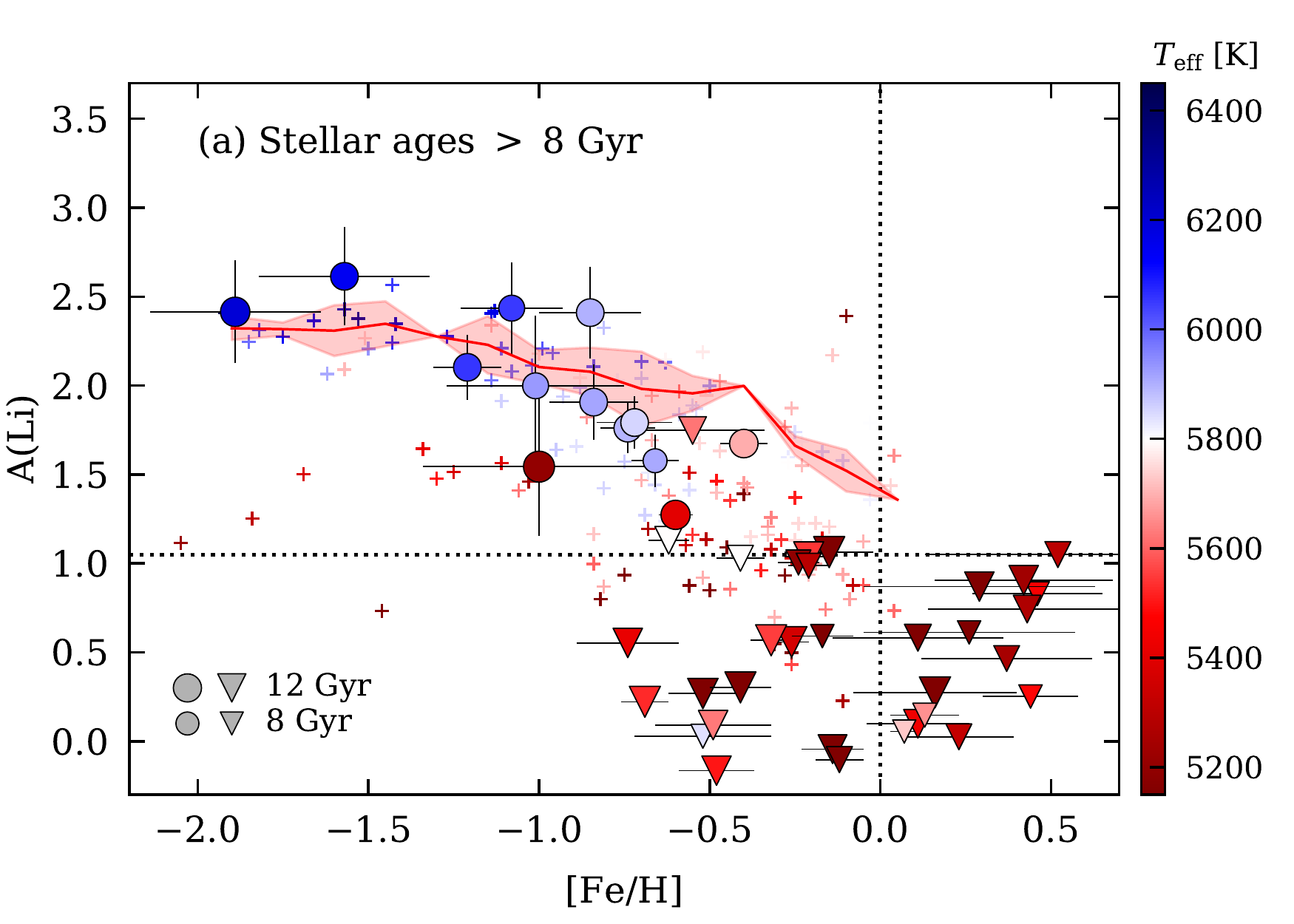}
\includegraphics[viewport= 0 0 504 350,clip]{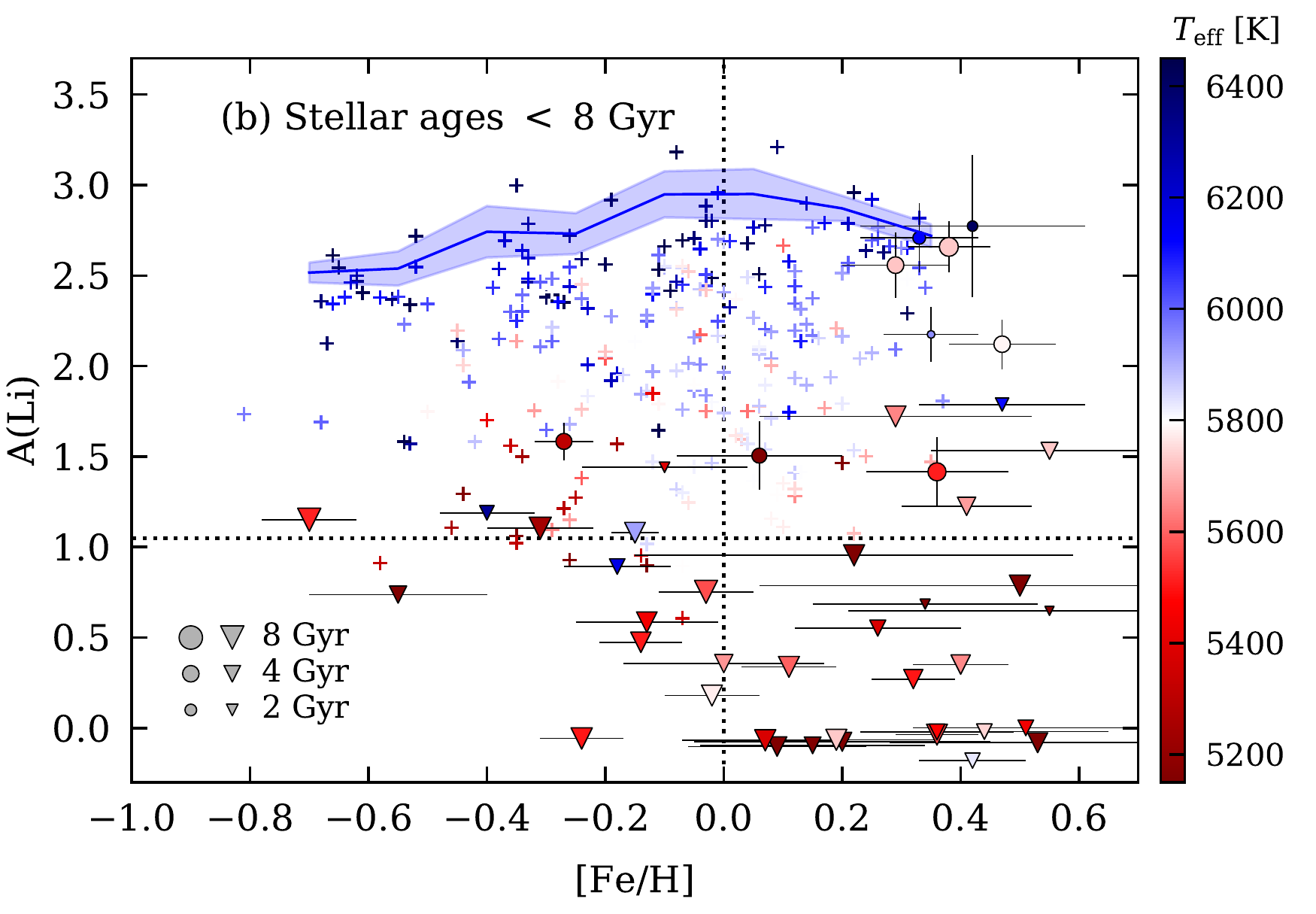}}
\caption{Li abundances versus [Fe/H] for the microlensed dwarf sample for stars older than 8\,Gyr (upper plot) and for stars younger than 8\,Gyr (lower plot). The stars have been colour-coded based on their effective temperatures (according to the colour-bar on the right-hand side). Triangles represent those stars where we could only determine upper limits on the abundances. The sizes of the circles and triangles are scaled by the ages of the stars, as shown in the lower-left corners of the plots. The coloured plus-signs in the background of each plot represent the stars from the sample of 714 stars from \cite{bensby2014} that are older and younger than 8\,Gyr, respectively. Li abundances were determined for these stars in \cite{bensby2018} in the exact same way as for the microlensed bulge dwarfs. The red and blue solid lines and the red and blue shaded regions show the running mean Li abundance, and the dispersion around the mean, calculated from the (up to) six stars with the highest Li abundances and that have $\teff>5800$\,K, in steps of 0.15\,dex and with a bin size of 0.2\,dex for the nearby disk stars. The dotted lines show the solar values with the solar photospheric Li abundance taken from \cite{asplund2009}.
\label{fig:li_feh_age}}
\end{figure*}

\subsection{Li, metallicity, and age}

Figure~\ref{fig:li_feh_age} shows again the Li abundances versus [Fe/H] plane for the microlensed bulge dwarfs, but now with the sample divided into a young subsample with stars younger than 8\,Gyr and an old subsample with stars older than 8\,Gyr. This division is inspired by the recent study of Li in the local disk by \cite{bensby2018} that shows two distinctly different Li trends by dividing the sample at 8\,Gyr. The old disk sample is representative of the thick disk and the young disk sample of the thin disk \citep[see also][]{bensby2014}. A separation between the thin and thick disks at about 8\,Gyr has been seen in other studies as well \citep[e.g.][]{haywood2013,bensby2014,kilic2017,silvaaguirre2018}.  

Figure~\ref{fig:li_feh_age} shows the running mean of the Li abundances for the disk stars, and the dispersion around the running mean, based on the (up to) six stars with the highest Li abundances and that have $\teff>5800$\,K. For the old disk sample the A(Li) trend is slightly decreasing with [Fe/H] (red line in Fig.~\ref{fig:li_feh_age}a), and for the young disk sample it is increasing with [Fe/H] until solar metallicities, after which it declines towards the highest metallicities (blue line in Fig.~\ref{fig:li_feh_age}b).

Within the uncertainties, the old bulge subsample with $\teff>5800$\,K follows the Li abundance pattern outlined by the old disk dwarfs in the Solar neighbourhood. The metal-poor stars tend to align with the Spite plateau \citep{spite1982}, and for the metal-rich part we see a declining trend as seen in the solar neighbourhood, meaning that no Li production took place during the first few billion years, coinciding with the era of the thick disk . One might argue that this declining A(Li) trend with [Fe/H] is an effect due to destruction of Li within the stars, as the stars at higher metallicities generally appear to have lower temperatures. However, within the small groups of stars hotter than 6000\,K, that should have not had any internal depletion, and between 5800-6000\, where Li might start to show some depletion, it is evident that the A(Li) trend is not increasing with metallicity.

For the bulge stars that are younger than 8\,Gyr and that have $\teff>5800$\,K we have three stars with well-determined Li abundances and five stars with upper limits. The two bulge stars with the highest Li abundances at $\rm [Fe/H]\approx +0.4$ have $\rm A(Li)\approx 2.8$, and seem to continue on the declining A(Li) trend with [Fe/H] as outlined by the disk stars.

\section{Discussion}

\begin{figure*}
\centering
\resizebox{0.75\hsize}{!}{
\includegraphics[viewport= 0 10 504 370,clip]{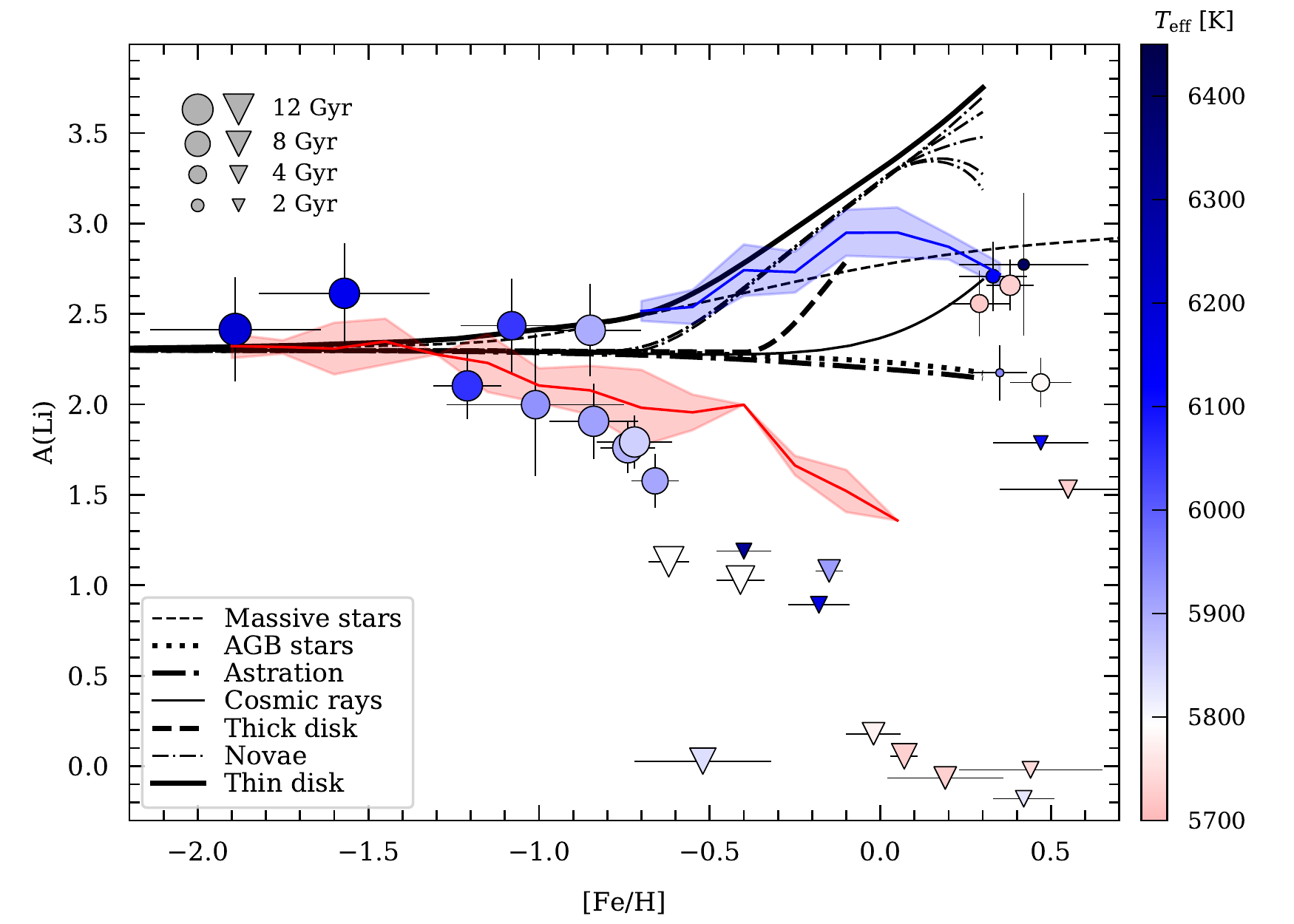}}
\caption{Comparison between the microlensed bulge dwarfs, the running mean A(Li) trends for the thin disk (younger than 8\,Gyr, blue line) and thick disk stars (older than 8\,Gyr, red line) from \cite{bensby2018}, and the different A(Li) evolution models from \cite{grisoni2019}.  The microlensed dwarf stars have been colour-coded with their effective temperatures (same scale as in previous figures) and their sizes have been scaled based on the estimated ages. The figure only includes bulge stars with effective temperatures greater than 5700\,K. The different chemical evolution models of Li are indicated in the legend.
\label{fig:li_feh_model}}
\end{figure*}

\subsection{Evolution of Li in the bulge}

\cite{grisoni2019} modelled the evolution of Li in the Milky Way and concluded that the main producer of Galactic Li is novae. By assuming that the fraction of binaries, which are supposed to be the novae progenitors, decrease with increasing metallicity, they also could provide an explanation for the decrease in A(Li) seen at the very highest metallicities. The thin dash-dotted lines in Fig.~\ref{fig:li_feh_model} shows the models from \cite{grisoni2019} where Li is produced by novae. There are several novae models shown, representing different laws for the fraction of binary systems that give rise to novae (see \citealt{grisoni2019} for more details). The solid thick line shows the total production of Li in the disk and includes other sources that are less significant, such as AGB stars (dotted line), cosmic rays (thin solid line), and massive stars (thin dashed line). The evolution of Li in the thick disk is shown by the thick dashed line, and the rise at solar metallicities is caused by Li production in novae that starts to contribute to the Galactic Li enrichment about a billion years after the beginning of Galaxy formation \citep{grisoni2019}. The thin dashed line shows the model for the Galactic bulge, which \cite{grisoni2019} assumes to be a classical bulge population, meaning no young stars, only an old classical spheroid. In that case the bulge evolved much faster than any of the disks, and long-lived stars did not have time to enrich the interstellar medium. For comparison purposes the thick dash-dotted line shows the Li evolution under pure astration, that is, no Li produced at all.

In Fig.~\ref{fig:li_feh_model} we compare the Li abundances for the microlensed bulge dwarfs to the models from \cite{grisoni2019}. To enhance the comparison we have excluded stars with $\teff<5700$\,K and as the colour-coding with temperature is the same as in the other figures, stars that have $\teff<5800$\,K are red in colour.  The figure also includes the running median Li abundances for the nearby thick disk stars (red line) and the nearby thin disk stars (blue line) from \cite{bensby2018}.

The A(Li)-[Fe/H] trend for the bulge stars older than about 8\,Gyr compares well with the trend for the local thick disk stars, and there is no indication of an increase in Li for the old bulge stars or for the nearby thick disk at higher metallicities.  The lack of bulge stars with well-determined Li abundances around solar metallicities makes it impossible to say if this truly is the case, or whether the thick disk model from \cite{grisoni2019} that shows the Li increase at $\rm [Fe/H]\approx-0.2$ (thick dashed line) is a proper representation. At the same time we see that there are a few stars at these metallicities, that have $\teff>5800$\,K, and for which only upper Li estimates could be given. Whether these stars were born with lower Li abundances or not remains unclear, but it is unlikely that they should have the high Li abundances corresponding to the thin disk (thick solid line) or thick disk (thick dashed line) models by \cite{grisoni2019}.

The few young (age less than about 8\,Gyr) bulge stars with $\teff>5700$\,K and that have Li abundance estimates are all located at very high metallicities, and most of them in such a way as  they appear to extend on the local thin thick disk Li trend. Interestingly, the same region is also the crossing area for the \cite{grisoni2019} models of Li enrichment from massive stars and cosmic rays. If we were to assume that the bulge is single classical bulge and the microlensed dwarf stars then all belonged to this population the match with the model showing the Li production by massive stars would striking. However, if this were the case, as the massive stars in the model contribute on a very short time scale, the metal-rich stars would have to be old, older than about 8 to 10 billion years as all the metal-poor bulge stars appear to be. This is clearly not the case, the metal-rich stars in Fig.~\ref{fig:li_feh_model} have ages below 4 billion years. So, we cannot make that connection or conclusion. Instead, if the bulge is a mixture of the other major stellar populations in the Milky Way, in particular the thin and thick disks, the the age and location of these few stars at the metal-rich end of the A(Li) trend connects them nicely to the thin disk. Also, their Li abundances seem to add further pieces of evidence of a declining A(Li) trend at super-solar metallicities. One should, however, caution that as we are lacking (young) stars with effective temperatures higher than 5800\,K around solar metallicities, this statement is not conclusive. Maybe it could also be that the Li abundances in the metal-rich bulge dwarfs hint at a different evolutionary path in the inner parts of the Milky Way, where novae has not played a major role, but instead cosmic rays has provided the necessary Li enrichment. More observations of relatively hot dwarf stars in the bulge is clearly needed to put this on firm ground.

\subsection{Radial migration of inner disk/bulge stars}

Recent developments in modelling of radial migration has indicated that it is the most metal-rich and kinematically cold stars that migrate the most. For instance, \cite{minchev2018} show that the stars in the metal-rich tail of the local metallicity distribution most likely were born at galactocentric radii smaller than about 4\,kpc. If the Li enrichment of the inner bulge region, or inner disk, has had a different Li enrichment than the rest of the Milky Way disk, and if some of those stars migrated to the Solar neighbourhood, could they be the answer to the decreasing Li trend at super-solar metallicities? This is the proposed solution by \cite{guiglion2019}, where they claim that the inner disk stars have experienced a faster enrichment and therefore a delayed Li production, meaning that the Li increase occurs at higher [Fe/H] than in the local disk. In fact, the metal-rich bulge stars in our sample do not reach as high Li abundances as seen in the local dwarf sample around $\rm [Fe/H]\approx0$, instead they cluster around $\rm A(Li) \approx 2.5-2.8$, appearing to extend on the declining A(Li) trend seen in the local disk stars (see Fig.~\ref{fig:li_feh_model}). Again, our sample does not contain any warm stars $\teff>5800$\,K with well-determined Li abundances around $\rm [Fe/H]\approx0$. That would have allowed us to distinguish between a continuously increasing Li trend in the bulge and a Li trend in the bulge that would have been similar to the local thin disk, that is, increasing towards solar metallicities and then declining. Currently we cannot rule out either of the two.  

We also note that \cite{pompeia2002} analysed Li in a sample of bulge-like dwarf stars, stars that have eccentric orbits with pericentric distances small as 2-3\,kpc, and found a large scatter in A(Li) and no apparent trend with [Fe/H]. 

\subsection{Similarities between the bulge and the local disk}

Several studies have claimed that the bulge, rather than being a single unique stellar population, contains contributions from the other main Galactic stellar populations (in particular the thin and thick disks) and that they happen to overlap in the bulge region \citep[e.g.][]{ness2013,bensby2017}. Apart from a multimodal metallicity distribution, the observed abundance trends in the bulge agree very well with the the abundance trends in the local thin and thick disks (see references in Sect.~\ref{sec:introduction}), which could point to possible connections between the populations and their evolutionary histories.

These similarities between the bulge and the local disks continue with Li. The observed A(Li) versus [Fe/H] trend in the (old) metal-poor bulge remains flat (or could even be slightly declining) and follows almost perfectly the observed trend in the local (old) thick disk. This indicates that no Li was produced during the first few billion years of the bulge era, or of the first few billion years of the thick disk era (as shown by \citealt{bensby2018}). The Li detections in microlensed dwarf sample at super-solar metallicities have poor statistics, but the few most metal-rich bulge stars Li abundances that extend on the decreasing Li trend seen in the local disk. We cannot, however, claim that this is truly the case, or whether the bulge has experienced a delayed Li enrichment due to faster chemical enrichment in the inner regions of the galaxy. It is clear, however, that the few metal-rich stars are young, and cannot be part of any classical bulge component.

\section{Summary}

We have determined Li abundances in 91 dwarf and subgiant stars in the Galactic bulge. For 22 of the stars the Li line at 6707.8\,{\AA} could be detected and Li abundances could be well determined. For the remaining 69 stars upper limits are for the Li abundance were estimated. The analysis was done by line synthesis of the $^7$Li line at 6707.8\,{\AA} in spectra that were obtained when the targets were microlensed and their brightnesses highly magnified. The methodology of determining stellar parameters and Li abundances is identical to ones in \cite{bensby2014} and \cite{bensby2018} where 714 local disk dwarf stars were analysed. This has enabled a truly differential comparison of the evolution of Li in the bulge and the local thin and thick disks. Our main findings are:

\begin{itemize}

\item The A(Li)-[Fe/H] trend for the bulge stars older than 8\,Gyr shows a very clear agreement to the A(Li)-[Fe/H] trend seen in the local thick disk (stars older than 8\,Gyr).

\item The A(Li)-[Fe/H] trend for the bulge stars older than 8\,Gyr remains flat, or might even slightly decline with increasing metallicity. This indicates that no Li was produced during the first few billion years of the bulge era, that happens to coincide with the first few billion years of the thick disk era (where no indications of Li production can be seen either).

\item The most metal-rich stars are young and have lithium abundances lower than the peak of the upper envelope of the Li distribution, that happens around $\rm [Fe/H]\approx 0$. They appear to extend on the decline in A(Li) at super-solar [Fe/H] that has been seen in recent studies of disk stars.

\item Whether the cause for the lower Li abundance in the most metal-rich bulge stars is due to the fact that they are the products of a similar enrichment history as seen in the local thin disk (with a maximum around solar metallicities and then a decline) or if they have followed another enrichment history, needs to be further investigated. The current sample is lacking stars around solar metallicities that have well-determined Li abundances. However, there are no indications that the lower Li abundances in the most metal-rich bulge stars is due to delayed Li production as a consequence of faster chemical enrichment in the bulge. Then these stars would have been much older than their current age estimates around 3-4 billion years

\end{itemize}

\begin{acknowledgement}

T.B. and S.F. were partly funded the project grant `The New Milky Way' project from the Knut and Alice Wallenberg Foundation. T.B. was partly funded by grant No. 2018-04857 from the Swedish Research Council. Work by J.C.Y. was supported by an SNSF Graduate Research Fellowship under Grant No. 2009068160. A.G. and J.C.Y. acknowledge support from NSF AST-1103471. M.A. gratefully acknowledges funding from the Australian Research Council (FL110100012). J.L.C. is grateful to NSF award AST-0908139 for partial support. S.L. research was partially supported by the DFG cluster of excellence `Origin and Structure of the Universe'. J.M. thanks support of FAPESP (2014/18100-4) and CNPq (Bolsa de Produtividade). We thank the referee, Piercarlo Bonifacio, for comments that helped to improve the paper. We also thank Valeria Grisoni for providing the Galactic chemical evolution models of Li. 
\end{acknowledgement}

\bibliographystyle{aa}
\bibliography{referenser}

\begin{thebibliography}{58}
\expandafter\ifx\csname natexlab\endcsname\relax\def\natexlab#1{#1}\fi

\bibitem[{{Alves-Brito} {et~al.}(2010){Alves-Brito}, {Mel{\'e}ndez}, {Asplund},
  {Ram{\'{\i}}rez}, \& {Yong}}]{alvesbrito2010}
{Alves-Brito}, A., {Mel{\'e}ndez}, J., {Asplund}, M., {Ram{\'{\i}}rez}, I., \&
  {Yong}, D. 2010, \aap, 513, A35

\bibitem[{{Asplund} {et~al.}(2009){Asplund}, {Grevesse}, {Sauval}, \&
  {Scott}}]{asplund2009}
{Asplund}, M., {Grevesse}, N., {Sauval}, A.~J., \& {Scott}, P. 2009, \araa, 47,
  481

\bibitem[{{Bensby} {et~al.}(2011){Bensby}, {Ad{\'e}n}, {Mel{\'e}ndez}, {Gould},
  {Feltzing}, {Asplund}, {Johnson}, {Lucatello}, {Yee}, {Ram{\'{\i}}rez},
  {Cohen}, {Thompson}, {Bond}, {Gal-Yam}, {Han}, {Sumi}, {Suzuki}, {Wada},
  {Miyake}, {Furusawa}, {Ohmori}, {Saito}, {Tristram}, \&
  {Bennett}}]{bensby2011}
{Bensby}, T., {Ad{\'e}n}, D., {Mel{\'e}ndez}, J., {et~al.} 2011, \aap, 533,
  A134

\bibitem[{{Bensby} {et~al.}(2010){Bensby}, {Asplund}, {Johnson}, {Feltzing},
  {Mel{\'e}ndez}, {Dong}, {Gould}, {Han}, {Ad{\'e}n}, {Lucatello}, \&
  {Gal-Yam}}]{bensby2010li}
{Bensby}, T., {Asplund}, M., {Johnson}, J.~A., {et~al.} 2010, \aap, 521, L57

\bibitem[{{Bensby} {et~al.}(2017){Bensby}, {Feltzing}, {Gould}, {Yee},
  {Johnson}, {Asplund}, {Mel{\'e}ndez}, {Lucatello}, {Howes}, {McWilliam},
  {Udalski}, {Szyma{\'n}ski}, {Soszy{\'n}ski}, {Poleski}, {Wyrzykowski},
  {Ulaczyk}, {Koz{\l}owski}, {Pietrukowicz}, {Skowron}, {Mr{\'o}z}, {Pawlak},
  {Abe}, {Asakura}, {Bhattacharya}, {Bond}, {Bennett}, {Hirao}, {Nagakane},
  {Koshimoto}, {Sumi}, {Suzuki}, \& {Tristram}}]{bensby2017}
{Bensby}, T., {Feltzing}, S., {Gould}, A., {et~al.} 2017, \aap, 605, A89

\bibitem[{{Bensby} {et~al.}(2014){Bensby}, {Feltzing}, \& {Oey}}]{bensby2014}
{Bensby}, T., {Feltzing}, S., \& {Oey}, M.~S. 2014, \aap, 562, A71

\bibitem[{{Bensby} \& {Lind}(2018)}]{bensby2018}
{Bensby}, T. \& {Lind}, K. 2018, \aap, 615, A151

\bibitem[{{Bensby} {et~al.}(2013){Bensby}, {Yee}, {Feltzing}, {Johnson},
  {Gould}, {Cohen}, {Asplund}, {Mel{\'e}ndez}, {Lucatello}, {Han}, {Thompson},
  {Gal-Yam}, {Udalski}, {Bennett}, {Bond}, {Kohei}, {Sumi}, {Suzuki}, {Suzuki},
  {Takino}, {Tristram}, {Yamai}, \& {Yonehara}}]{bensby2013}
{Bensby}, T., {Yee}, J.~C., {Feltzing}, S., {et~al.} 2013, \aap, 549, A147

\bibitem[{{Boesgaard} {et~al.}(2016){Boesgaard}, {Lum}, {Deliyannis}, {King},
  {Pinsonneault}, \& {Somers}}]{boesgaard2016}
{Boesgaard}, A.~M., {Lum}, M.~G., {Deliyannis}, C.~P., {et~al.} 2016, \apj,
  830, 49

\bibitem[{{Boesgaard} \& {Tripicco}(1986)}]{boesgaard1986}
{Boesgaard}, A.~M. \& {Tripicco}, M.~J. 1986, \apjl, 302, L49

\bibitem[{{Bonifacio} {et~al.}(2019){Bonifacio}, {Caffau}, {Spite}, \&
  {Spite}}]{bonifacio2019}
{Bonifacio}, P., {Caffau}, E., {Spite}, M., \& {Spite}, F. 2019, Research Notes
  of the American Astronomical Society, 3, 64

\bibitem[{{Bonifacio} {et~al.}(2018){Bonifacio}, {Caffau}, {Spite}, {Spite},
  {Sbordone}, {Monaco}, {Fran{\c c}ois}, {Plez}, {Molaro}, {Gallagher},
  {Cayrel}, {Christlieb}, {Klessen}, {Koch}, {Ludwig}, {Steffen}, {Zaggia}, \&
  {Abate}}]{bonifacio2018}
{Bonifacio}, P., {Caffau}, E., {Spite}, M., {et~al.} 2018, \aap, 612, A65

\bibitem[{{Buder} {et~al.}(2018){Buder}, {Asplund}, {Duong}, {Kos}, {Lind},
  {Ness}, {Sharma}, {Bland-Hawthorn}, {Casey}, {De Silva}, {D'Orazi},
  {Freeman}, {Lewis}, {Lin}, {Martell}, {Schlesinger}, {Simpson}, {Zucker},
  {Zwitter}, {Amarsi}, {Anguiano}, {Carollo}, {Casagrande}, {{\v C}otar},
  {Cottrell}, {Da Costa}, {Gao}, {Hayden}, {Horner}, {Ireland}, {Kafle},
  {Munari}, {Nataf}, {Nordlander}, {Stello}, {Ting}, {Traven}, {Watson},
  {Wittenmyer}, {Wyse}, {Yong}, {Zinn}, \& {{\v Z}erjal}}]{buder2018}
{Buder}, S., {Asplund}, M., {Duong}, L., {et~al.} 2018, \mnras, 478, 4513

\bibitem[{{Carney} {et~al.}(2005){Carney}, {Latham}, \& {Laird}}]{carney2005b}
{Carney}, B.~W., {Latham}, D.~W., \& {Laird}, J.~B. 2005, \aj, 129, 466

\bibitem[{{Cavallo} {et~al.}(2003){Cavallo}, {Cook}, {Minniti}, \&
  {Vandehei}}]{cavallo2003}
{Cavallo}, R.~M., {Cook}, K.~H., {Minniti}, D., \& {Vandehei}, T. 2003, in
  Society of Photo-Optical Instrumentation Engineers (SPIE) Conference Series,
  Vol. 4834, Society of Photo-Optical Instrumentation Engineers (SPIE)
  Conference Series, ed. P.~{Guhathakurta}, 66--73

\bibitem[{{Cescutti} \& {Molaro}(2019)}]{cescutti2019}
{Cescutti}, G. \& {Molaro}, P. 2019, \mnras, 482, 4372

\bibitem[{{Chen} {et~al.}(2001){Chen}, {Nissen}, {Benoni}, \&
  {Zhao}}]{chen2001}
{Chen}, Y.~Q., {Nissen}, P.~E., {Benoni}, T., \& {Zhao}, G. 2001, \aap, 371,
  943

\bibitem[{{Cyburt} {et~al.}(2016){Cyburt}, {Fields}, {Olive}, \&
  {Yeh}}]{cyburt2016}
{Cyburt}, R.~H., {Fields}, B.~D., {Olive}, K.~A., \& {Yeh}, T.-H. 2016, Reviews
  of Modern Physics, 88, 015004

\bibitem[{{D'Antona} \& {Matteucci}(1991)}]{dantona1991}
{D'Antona}, F. \& {Matteucci}, F. 1991, \aap, 248, 62

\bibitem[{{De Silva} {et~al.}(2015){De Silva}, {Freeman}, {Bland-Hawthorn},
  {Martell}, {de Boer}, {Asplund}, {Keller}, {Sharma}, {Zucker}, {Zwitter},
  {Anguiano}, {Bacigalupo}, {Bayliss}, {Beavis}, {Bergemann}, {Campbell},
  {Cannon}, {Carollo}, {Casagrande}, {Casey}, {Da Costa}, {D'Orazi}, {Dotter},
  {Duong}, {Heger}, {Ireland}, {Kafle}, {Kos}, {Lattanzio}, {Lewis}, {Lin},
  {Lind}, {Munari}, {Nataf}, {O'Toole}, {Parker}, {Reid}, {Schlesinger},
  {Sheinis}, {Simpson}, {Stello}, {Ting}, {Traven}, {Watson}, {Wittenmyer},
  {Yong}, \& {{\v Z}erjal}}]{desilva2015}
{De Silva}, G.~M., {Freeman}, K.~C., {Bland-Hawthorn}, J., {et~al.} 2015,
  \mnras, 449, 2604

\bibitem[{{Delgado Mena} {et~al.}(2015){Delgado Mena}, {Bertr{\'a}n de Lis},
  {Adibekyan}, {Sousa}, {Figueira}, {Mortier}, {Gonz{\'a}lez Hern{\'a}ndez},
  {Tsantaki}, {Israelian}, \& {Santos}}]{delgadomena2015}
{Delgado Mena}, E., {Bertr{\'a}n de Lis}, S., {Adibekyan}, V.~Z., {et~al.}
  2015, \aap, 576, A69

\bibitem[{{Delgado Mena} {et~al.}(2014){Delgado Mena}, {Israelian},
  {Gonz{\'a}lez Hern{\'a}ndez}, {Sousa}, {Mortier}, {Santos}, {Adibekyan},
  {Fernandes}, {Rebolo}, {Udry}, \& {Mayor}}]{delgadomena2014}
{Delgado Mena}, E., {Israelian}, G., {Gonz{\'a}lez Hern{\'a}ndez}, J.~I.,
  {et~al.} 2014, \aap, 562, A92

\bibitem[{{Fu} {et~al.}(2018){Fu}, {Romano}, {Bragaglia}, {Mucciarelli},
  {Lind}, {Delgado Mena}, {Sousa}, {Randich}, {Bressan}, {Sbordone}, {Martell},
  {Korn}, {Abia}, {Smiljanic}, {Jofr{\'e}}, {Pancino}, {Tautvai{\v s}ien{\.e}},
  {Tang}, {Magrini}, {Lanzafame}, {Carraro}, {Bensby}, {Damiani}, {Alfaro},
  {Flaccomio}, {Morbidelli}, {Zaggia}, {Lardo}, {Monaco}, {Frasca}, {Donati},
  {Drazdauskas}, {Chorniy}, {Bayo}, \& {Kordopatis}}]{fu2018}
{Fu}, X., {Romano}, D., {Bragaglia}, A., {et~al.} 2018, \aap, 610, A38

\bibitem[{{Ghezzi} {et~al.}(2010){Ghezzi}, {Cunha}, {Smith}, \& {de la
  Reza}}]{ghezzi2010li}
{Ghezzi}, L., {Cunha}, K., {Smith}, V.~V., \& {de la Reza}, R. 2010, \apj, 724,
  154

\bibitem[{{Gilmore} {et~al.}(2012){Gilmore}, {Randich}, {Asplund}, {Binney},
  {Bonifacio}, {Drew}, {Feltzing}, {Ferguson}, {Jeffries}, {Micela},
  {Negueruela}, {Prusti}, {Rix}, {Vallenari}, {Alfaro}, {Allende-Prieto},
  {Babusiaux}, {Bensby}, {Blomme}, {Bragaglia}, {Flaccomio}, {Francois},
  {Irwin}, {Koposov}, {Korn}, {Lanzafame}, {Pancino}, {Paunzen},
  {Recio-Blanco}, {Sacco}, {Smiljanic}, {van Eck}, \& {Walton}}]{gilmore2012}
{Gilmore}, G., {Randich}, S., {Asplund}, M., {et~al.} 2012, The Messenger, 147,
  25

\bibitem[{{Glaspey} {et~al.}(1994){Glaspey}, {Pritchet}, \&
  {Stetson}}]{glaspey1994}
{Glaspey}, J.~W., {Pritchet}, C.~J., \& {Stetson}, P.~B. 1994, \aj, 108, 271

\bibitem[{{Glebbeek} {et~al.}(2010){Glebbeek}, {Sills}, {Hu}, \&
  {Stancliffe}}]{glebbeek2010}
{Glebbeek}, E., {Sills}, A., {Hu}, H., \& {Stancliffe}, R.~J. 2010, in American
  Institute of Physics Conference Series, Vol. 1314, American Institute of
  Physics Conference Series, ed. V.~{Kalogera} \& M.~{van der Sluys}, 113--119

\bibitem[{{Gonzalez} {et~al.}(2011){Gonzalez}, {Rejkuba}, {Zoccali}, {Hill},
  {Battaglia}, {Babusiaux}, {Minniti}, {Barbuy}, {Alves-Brito}, {Renzini},
  {Gomez}, \& {Ortolani}}]{gonzalez2011}
{Gonzalez}, O.~A., {Rejkuba}, M., {Zoccali}, M., {et~al.} 2011, \aap, 530, A54

\bibitem[{{Gonzalez} {et~al.}(2009){Gonzalez}, {Zoccali}, {Monaco}, {Hill},
  {Cassisi}, {Minniti}, {Renzini}, {Barbuy}, {Ortolani}, \&
  {Gomez}}]{gonzalez2009}
{Gonzalez}, O.~A., {Zoccali}, M., {Monaco}, L., {et~al.} 2009, \aap, 508, 289

\bibitem[{{Grisoni} {et~al.}(2019){Grisoni}, {Matteucci}, {Romano}, \&
  {Fu}}]{grisoni2019}
{Grisoni}, V., {Matteucci}, F., {Romano}, D., \& {Fu}, X. 2019, \mnras, 489,
  3539

\bibitem[{{Guiglion} {et~al.}(2019){Guiglion}, {Chiappini}, {Romano},
  {Matteucci}, {Anders}, {Steinmetz}, {Minchev}, {de Laverny}, \&
  {Recio-Blanco}}]{guiglion2019}
{Guiglion}, G., {Chiappini}, C., {Romano}, D., {et~al.} 2019, \aap, 623, A99

\bibitem[{{Guiglion} {et~al.}(2016){Guiglion}, {de Laverny}, {Recio-Blanco},
  {Worley}, {De Pascale}, {Masseron}, {Prantzos}, \&
  {Mikolaitis}}]{guiglion2016}
{Guiglion}, G., {de Laverny}, P., {Recio-Blanco}, A., {et~al.} 2016, \aap, 595,
  A18

\bibitem[{{Gustafsson} {et~al.}(2008){Gustafsson}, {Edvardsson}, {Eriksson},
  {J{\o}rgensen}, {Nordlund}, \& {Plez}}]{gustafsson2008}
{Gustafsson}, B., {Edvardsson}, B., {Eriksson}, K., {et~al.} 2008, \aap, 486,
  951

\bibitem[{{Haywood} {et~al.}(2013){Haywood}, {Di Matteo}, {Lehnert}, {Katz}, \&
  {G{\'o}mez}}]{haywood2013}
{Haywood}, M., {Di Matteo}, P., {Lehnert}, M.~D., {Katz}, D., \& {G{\'o}mez},
  A. 2013, \aap, 560, A109

\bibitem[{{J{\"o}nsson} {et~al.}(2017){J{\"o}nsson}, {Ryde}, {Schultheis}, \&
  {Zoccali}}]{jonsson2017}
{J{\"o}nsson}, H., {Ryde}, N., {Schultheis}, M., \& {Zoccali}, M. 2017, \aap,
  598, A101

\bibitem[{{Kilic} {et~al.}(2017){Kilic}, {Munn}, {Harris}, {von Hippel},
  {Liebert}, {Williams}, {Jeffery}, \& {DeGennaro}}]{kilic2017}
{Kilic}, M., {Munn}, J.~A., {Harris}, H.~C., {et~al.} 2017, \apj, 837, 162

\bibitem[{{Lambert} \& {Reddy}(2004)}]{lambert2004}
{Lambert}, D.~L. \& {Reddy}, B.~E. 2004, \mnras, 349, 757

\bibitem[{{Lebzelter} {et~al.}(2012){Lebzelter}, {Uttenthaler}, {Busso},
  {Schultheis}, \& {Aringer}}]{lebzelter2012li}
{Lebzelter}, T., {Uttenthaler}, S., {Busso}, M., {Schultheis}, M., \&
  {Aringer}, B. 2012, \aap, 538, A36

\bibitem[{{Lind} {et~al.}(2009){Lind}, {Asplund}, \& {Barklem}}]{lind2009}
{Lind}, K., {Asplund}, M., \& {Barklem}, P.~S. 2009, \aap, 503, 541

\bibitem[{{Matteucci}(2010)}]{matteucci2010}
{Matteucci}, F. 2010, in IAU Symposium, Vol. 268, Light Elements in the
  Universe, ed. C.~{Charbonnel}, M.~{Tosi}, F.~{Primas}, \& C.~{Chiappini},
  453--461

\bibitem[{{Mel{\'e}ndez} {et~al.}(2008){Mel{\'e}ndez}, {Asplund},
  {Alves-Brito}, {Cunha}, {Barbuy}, {Bessell}, {Chiappini}, {Freeman},
  {Ram{\'{\i}}rez}, {Smith}, \& {Yong}}]{melendez2008}
{Mel{\'e}ndez}, J., {Asplund}, M., {Alves-Brito}, A., {et~al.} 2008, \aap, 484,
  L21

\bibitem[{{Mel{\'e}ndez} {et~al.}(2010){Mel{\'e}ndez}, {Casagrande},
  {Ram{\'{\i}}rez}, {Asplund}, \& {Schuster}}]{melendez2010li}
{Mel{\'e}ndez}, J., {Casagrande}, L., {Ram{\'{\i}}rez}, I., {Asplund}, M., \&
  {Schuster}, W.~J. 2010, \aap, 515, L3

\bibitem[{{Minchev} {et~al.}(2018){Minchev}, {Anders}, {Recio-Blanco},
  {Chiappini}, {de Laverny}, {Queiroz}, {Steinmetz}, {Adibekyan}, {Carrillo},
  {Cescutti}, {Guiglion}, {Hayden}, {de Jong}, {Kordopatis}, {Majewski},
  {Martig}, \& {Santiago}}]{minchev2018}
{Minchev}, I., {Anders}, F., {Recio-Blanco}, A., {et~al.} 2018, \mnras, 481,
  1645

\bibitem[{{Minniti} {et~al.}(1998){Minniti}, {Vandehei}, {Cook}, {Griest}, \&
  {Alcock}}]{minniti1998}
{Minniti}, D., {Vandehei}, T., {Cook}, K.~H., {Griest}, K., \& {Alcock}, C.
  1998, \apjl, 499, L175

\bibitem[{{Ness} {et~al.}(2013){Ness}, {Freeman}, {Athanassoula},
  {Wylie-de-Boer}, {Bland-Hawthorn}, {Asplund}, {Lewis}, {Yong}, {Lane}, \&
  {Kiss}}]{ness2013}
{Ness}, M., {Freeman}, K., {Athanassoula}, E., {et~al.} 2013, \mnras, 430, 836

\bibitem[{{Pavlenko} {et~al.}(2018){Pavlenko}, {Jenkins}, {Ivanyuk}, {Jones},
  {Kaminsky}, {Lyubchik}, \& {Yakovina}}]{pavlenko2018}
{Pavlenko}, Y.~V., {Jenkins}, J.~S., {Ivanyuk}, O.~M., {et~al.} 2018, \aap,
  611, A27

\bibitem[{{Pomp{\'e}ia} {et~al.}(2002){Pomp{\'e}ia}, {Barbuy}, {Grenon}, \&
  {Castilho}}]{pompeia2002}
{Pomp{\'e}ia}, L., {Barbuy}, B., {Grenon}, M., \& {Castilho}, B.~V. 2002, \apj,
  570, 820

\bibitem[{{Prantzos} {et~al.}(2017){Prantzos}, {de Laverny}, {Guiglion},
  {Recio-Blanco}, \& {Worley}}]{prantzos2017}
{Prantzos}, N., {de Laverny}, P., {Guiglion}, G., {Recio-Blanco}, A., \&
  {Worley}, C.~C. 2017, \aap, 606, A132

\bibitem[{{Ram{\'{\i}}rez} {et~al.}(2012){Ram{\'{\i}}rez}, {Fish}, {Lambert},
  \& {Allende Prieto}}]{ramirez2012}
{Ram{\'{\i}}rez}, I., {Fish}, J.~R., {Lambert}, D.~L., \& {Allende Prieto}, C.
  2012, \apj, 756, 46

\bibitem[{{Romano} {et~al.}(1999){Romano}, {Matteucci}, {Molaro}, \&
  {Bonifacio}}]{romano1999}
{Romano}, D., {Matteucci}, F., {Molaro}, P., \& {Bonifacio}, P. 1999, \aap,
  352, 117

\bibitem[{{Romano} {et~al.}(2001){Romano}, {Matteucci}, {Ventura}, \&
  {D'Antona}}]{romano2001}
{Romano}, D., {Matteucci}, F., {Ventura}, P., \& {D'Antona}, F. 2001, \aap,
  374, 646

\bibitem[{{Ryan} {et~al.}(2001){Ryan}, {Beers}, {Kajino}, \&
  {Rosolankova}}]{ryan2001b}
{Ryan}, S.~G., {Beers}, T.~C., {Kajino}, T., \& {Rosolankova}, K. 2001, \apj,
  547, 231

\bibitem[{{Sbordone} {et~al.}(2010){Sbordone}, {Bonifacio}, {Caffau}, {Ludwig},
  {Behara}, {Gonz{\'a}lez Hern{\'a}ndez}, {Steffen}, {Cayrel}, {Freytag},
  {van't Veer}, {Molaro}, {Plez}, {Sivarani}, {Spite}, {Spite}, {Beers},
  {Christlieb}, {Fran{\c c}ois}, \& {Hill}}]{sbordone2010}
{Sbordone}, L., {Bonifacio}, P., {Caffau}, E., {et~al.} 2010, \aap, 522, A26

\bibitem[{{Schirbel} {et~al.}(2015){Schirbel}, {Mel{\'e}ndez}, {Karakas},
  {Ram{\'\i}rez}, {Castro}, {Faria}, {Lugaro}, {Asplund}, {Tucci Maia}, {Yong},
  {Howes}, \& {do Nascimento}}]{schirbel2015}
{Schirbel}, L., {Mel{\'e}ndez}, J., {Karakas}, A.~I., {et~al.} 2015, \aap, 584,
  A116

\bibitem[{{Silva Aguirre} {et~al.}(2018){Silva Aguirre}, {Bojsen-Hansen},
  {Slumstrup}, {Casagrande}, {Kawata}, {Ciuc{\v a}}, {Handberg}, {Lund},
  {Mosumgaard}, {Huber}, {Johnson}, {Pinsonneault}, {Serenelli}, {Stello},
  {Tayar}, {Bird}, {Cassisi}, {Hon}, {Martig}, {Nissen}, {Rix},
  {Sch{\"o}nrich}, {Sahlholdt}, {Trick}, \& {Yu}}]{silvaaguirre2018}
{Silva Aguirre}, V., {Bojsen-Hansen}, M., {Slumstrup}, D., {et~al.} 2018,
  \mnras, 475, 5487

\bibitem[{{Spite} \& {Spite}(1982)}]{spite1982}
{Spite}, F. \& {Spite}, M. 1982, \aap, 115, 357

\bibitem[{{Uttenthaler} {et~al.}(2007){Uttenthaler}, {Lebzelter}, {Palmerini},
  {Busso}, {Aringer}, \& {Lederer}}]{uttenthaler2007li}
{Uttenthaler}, S., {Lebzelter}, T., {Palmerini}, S., {et~al.} 2007, \aap, 471,
  L41

\bibitem[{{Valenti} \& {Piskunov}(1996)}]{valenti1996}
{Valenti}, J.~A. \& {Piskunov}, N. 1996, \aaps, 118, 595

\end{thebibliography}

\end{document}